\documentclass[12pt]{amsart}

\usepackage{subfigure}
\usepackage{natbib}
\usepackage{url}

\usepackage{amsaddr}
\usepackage{float}
\usepackage{hyperref}
\usepackage[pdftex]{graphicx}
\usepackage{caption}
\usepackage[a4paper, margin=2cm]{geometry}
\usepackage{booktabs}

\theoremstyle{plain}

\theoremstyle{definition}

\theoremstyle{remark}

\begin{document}

\title{Predicting school transition rates in Austria with classification trees}

\author{Annette M\"{o}ller}
\address{Faculty of Business Administration and Economics,  Bielefeld University, Germany}
\email{annette.moeller@uni-bielefeld.de}
\author{Ann Cathrice George}
\address{Federal Institute for Quality Assurance of the Austrian School System (IQS), Austria}
\email{anncathrice.george@iqs.gv.at}
\author{J\"{u}rgen Gro{\ss}}
\address{Institute for Mathematics and Applied Informatics, University of Hildesheim, Germany}
\email{juergen.gross@uni-hildesheim.de}
\thanks{This is an original manuscript of an article published by Taylor \& Francis in International
Journal of Research \& Method on 17 Oct 2022, available at: http://www.tandfonline.com/doi/full/10.1080/1743727X.2022.2128744.}

\begin{abstract}
Methods based on machine learning become increasingly popular in many areas as they allow models to be fitted in a highly-data driven fashion, and often show comparable or even increased performance in comparison to classical methods.
However, in the area of educational sciences the application of machine learning is still quite uncommon.
This work investigates the benefit of using classification trees for analyzing data from educational sciences. An application to data on school transition rates in Austria indicates different aspects of interest in the context of educational sciences: (i) the trees select variables for predicting school transition rates in a data-driven fashion which are well in accordance with existing confirmatory theories from educational sciences, (ii) trees can be employed for performing variable selection for regression models, (iii) the classification performance of trees is comparable to that of binary regression models.
These results indicate that trees and possibly other machine learning methods may also be helpful to explore high-dimensional educational data sets, especially where no confirmatory theories have been developed yet.
\end{abstract}


\maketitle

\section{Introduction} \label{sec:Intro}

Machine learning methods become more and more popular in many applications and often show competitive performance to traditional models from applied statistics, as e.g. regression models. Regression and classification trees have many appealing advantages. In contrast to traditional regression models they can deal with very large numbers of predictors and do not require any assumptions regarding distribution or the relationship of predictors and response. Furthermore, a tree model is highly interpretable due to its hierarchical nature.

Although the use of machine learning is still rare in educational sciences, it recently becomes evident that applying these methods in combination with traditional statistical methods is helpful in many ways \citep{LezhninaKismihok2021}.
A few approaches utilizing machine learning are given in the following: In \cite{Sinharay2016} trees, random forests and boosting are employed to predict different variables of educational interest, as for example
item difficulty, high school dropouts, and scorings in electronic essays. The authors note that these methods have slightly superior performance over traditional regression models. The work in \cite{GaoRogers2011} uses regression trees to predict and interpret item difficulties in a language assessment survey and concluded that the tree structure can be used to enhance interpretation of the items. The authors in \cite{Salles&2020} made attempts to analyze data from computer based assessments, which pose challenges to the researcher due to their high-dimensionality.

In order to continue and extend the research conducted so far in educational sciences the issues (i) - (iii) stated in the following will be addressed in the subsequent analysis:

\noindent
(i) Are the predictors selected in the tree in accordance with existing theories from educational sciences or can even complement them (Section \ref{sec:relevancepredictors})? \\
(ii) Is it possible to utilize the variable choices of a tree as data-driven variable selection method for building regression models to predict an educational response variable? Can such a tree-based variable selection be supportive and complementary when choosing variables based on educational approaches (Section \ref{sec:variablesel})? \\
(iii) Is the performance of a tree comparable to the performance of traditional generalized linear (mixed) models when applied to educational data (Section \ref{sec:predperf})?

In this regard classification trees are employed to predict school transition rates in Austria based on data of the test of educational standards in mathematics for fourth graders \citep{BIFIE2019}.

If the variable selection of trees indeed leads to reasonable interpretations, future research may consider the application of trees or other machine learning methods to data and research questions in which no educational theories exist so far. This possibility is further outlined in Section \ref{sec:outlook}.

\section{Background information on data and investigated educational topic} \label{sec:data}

\subsection{Educational research topic and existing educational approaches}

The Organization for Cooperation and Development \citep{OECD2010} has shown that unemployment rates decrease as the level of education raises. Understanding the factors related to educational aspiration is of great interest to educators in order to explain and predict the choices students make during their educational paths. Students in Austria make the first choice regarding their educational path after grade four: for the upcoming school transition they can select between a higher academic track, called ``Allgemeine H\"{o}here Schule'' (AHS), which (after graduation) allows them to enroll at a university, and lower academic tracks. Insights from educational science show that not only the students' competencies are of relevance for that selection but also other social and personal factors.
The present study follows the (existing) approach of \cite{Flores&2011} which model students' aspiration in dependence of a small number of variables they indicated in a comprehensive literature review. These variables are gender, educational status of parents, educational attainment, socio-economic status (measured by the number of books in the household), and parental involvement in school (measured by the students' equipment regarding a working place at home).

\subsection{Background on data and variables}

Information about students' educational aspiration, i.e. their plans which school to visit after grade four are given in the data of the Austrian educational standard tests for fourth graders.
The Austrian educational standard tests are employed to monitor students' competencies, and based on the results, to enhance the school system. This study is based on data of the Austrian educational standard test in mathematics for fourth graders in 2018 \citep{BIFIE2019}. The Austrian standards testing is mandatory, which leads to a survey of 73,780 students in 4,925 classes and 2,961 schools. The present study includes a representative sample of 8,520 students in 637 classes and 430 schools.

The data mainly includes students' competencies in mathematics which can be categorized into four content sub-domains (i.e. numbers, operations, measures, geometry), and four cognitive sub-domains (i.e. model building, calculating, communicating, problem solving), with each of them measured in the test
(for measuring students' competencies in all eight domains see also \citealp{Gross&2016}).
The results are given on continuous scales with a mean value of 500.

Besides the students' competencies additional background information of students, teachers, parents and schools is collected via context questionnaires \citep{BIFIE2018}. The questionnaires yield multifaceted information about the students' family and school environment, as well as personal and motivational factors.
As mentioned above, the variable to be predicted in this study is the school the students will attend in the following year.

\begin{table}
\caption{Overview on important variables in data set. The first column contains the variable names, the second column states the level on which a variable is measured (in brackets the level of the corresponding aggregated variable), the third column provides a short explanation of each variable.}
{\begin{tabular}{lll} \toprule
Variable & Data Level & Description \\
\midrule
\begin{minipage}{0.3\textwidth} \texttt{aspiration-education}\end{minipage} &
\begin{minipage}{0.1\textwidth}
student\end{minipage} & \begin{minipage}{0.5\textwidth} Parents aspiration of the highest education their children will achieve (higher values indicate higher education)\end{minipage} \\
\midrule
\begin{minipage}{0.3\textwidth} \texttt{math-grade}\end{minipage} & \begin{minipage}{0.1\textwidth} student
\end{minipage} & \begin{minipage}{0.5\textwidth}Grade students achieve in mathematics (higher values indicate poorer grades)\end{minipage} \\
\midrule
\begin{minipage}{0.3\textwidth}
 \texttt{points-calculating}\end{minipage} & \begin{minipage}{0.1\textwidth}
 student\end{minipage} & \begin{minipage}{0.5\textwidth} Points in the educational standards test in cognitive sub-domain calculating  (higher values for higher competencies)\end{minipage}\\
 \midrule
\begin{minipage}{0.3\textwidth}
 \texttt{points-communicating}\end{minipage} & \begin{minipage}{0.1\textwidth}student\end{minipage} & \begin{minipage}{0.5\textwidth} Points in the educational standards test in cognitive sub-domain communicating about mathematical facts
 (higher values for higher competencies)\end{minipage}\\
 \midrule
 \begin{minipage}{0.3\textwidth}
 \texttt{private-tutoring}\\
  (\texttt{private-tutoring-aggCL})
  \end{minipage} &
  \begin{minipage}{0.1\textwidth} student\\
  (class)\end{minipage}
   & \begin{minipage}{0.5\textwidth} Number of hours for private tutoring in class
  (categorized, 7 levels, higher values indicate more hours)
  \end{minipage} \\
\midrule
\begin{minipage}{0.3\textwidth}
 \texttt{after-school}\\
 (\texttt{after-school-aggSL})\end{minipage} &
 \begin{minipage}{0.1\textwidth} student\\(school)
 \end{minipage}
 &
 \begin{minipage}{0.5\textwidth}
  Visiting after-school supervision (higher values indicate more time spent in supervision)
  \end{minipage}\\
\midrule
\begin{minipage}{0.3\textwidth}
 \texttt{social-status}\\
 (\texttt{social-status-aggSL})
 \end{minipage} & \begin{minipage}{0.1\textwidth} student\\
 (school)\end{minipage} & \begin{minipage}{0.5\textwidth}A measure indicating the social status of the student (higher values indicate higher social status)\end{minipage}\\
 \midrule
\begin{minipage}{0.3\textwidth}
 \texttt{federal-state}\end{minipage} & \begin{minipage}{0.1\textwidth}school\end{minipage}
  & \begin{minipage}{0.5\textwidth} Federal state of school location (9 states in Austria)\end{minipage}\\
\midrule
\begin{minipage}{0.3\textwidth}
 \texttt{town-size}\end{minipage} & \begin{minipage}{0.1\textwidth} school\end{minipage}
  & \begin{minipage}{0.5\textwidth} Size of the town where the school is located (higher values indicate larger towns)
   \end{minipage}\\
\midrule
\begin{minipage}{0.3\textwidth}
 \texttt{school-size}
 \end{minipage}& \begin{minipage}{0.1\textwidth} school\end{minipage} & \begin{minipage}{0.5\textwidth}
 Total number of students attending the school\end{minipage}\\
\midrule
\begin{minipage}{0.3\textwidth}
 \texttt{urban}\end{minipage} & \begin{minipage}{0.1\textwidth} school\end{minipage} &
 \begin{minipage}{0.5\textwidth} Degree of urbanization where the school is located (higher values indicate a smaller degree of urbanization)\end{minipage}\\
\bottomrule
\end{tabular}}
\label{InfoVariables}
\end{table}

The around 700 context variables obtained from the complete questionnaires will be employed in the subsequent analysis to predict school transition.
Table \ref{InfoVariables} presents a selection of context variables, which will be found to be of relevance in the subsequent model building process, see, e.g., Section \ref{sec:relevancepredictors}.

\subsection{Pre-processing of data} \label{sec:preproc}

The variables appearing in the data set are measured on different scale levels. To account for the nature of the variables correctly, appropriate coding was conducted beforehand. Several variables were coded as factor (categorical variables on nominal scale) or as ordered factor (categorical variables on ordinal scale).

\subsubsection{Merging questionnaires}

Some of the context variables appear in the parents' as well as in the students' questionnaire. For example parents and students were asked about the (categorized) number of books in their household. In such cases the information was summarized into a single variable in the following way: if the parents' questionnaire provided an answer (this questionnaire was voluntary), it was taken for the variables entry, regardless of the students' answer. If the parents' answer was missing and the students' answer given, the students' answer was taken. In case of a missing value in both questionnaires the variables entry was left to missing. This procedure was also applied to the dependent (response) variable of school selection.
Both, students and parents, could select between four options: ``Neue Mittelschule'' (describing a lower academic track), ``AHS'' (the higher academic track), ``another type of school'' (i. e. other secondary schools or special schools) or ``I don’t know''.
Since parents must register their children for the next higher school in January the standards tests and questionnaires are performed
in April. As students will change schools after vacations in August the answers to that question can be expected to be of high precision.
For this analysis the students' and parents' answers were combined and the result dichotomized.
The option ``AHS'' was coded as \texttt{1} and all other options as \texttt{0}, leading to a binary response variable used in the modelling process.
As a result, there are still missing values in several of the variables, only the response variable was recoded to contain no missing values.
Of all students in the analyzed data set $\approx$ 40 \% indicated to visit AHS in the following year.

\subsubsection{Multilevel structure of data} \label{sec:multilevel}

The data set contains context variables collected on individual/student level, class level and school level.
The tree models considered here have no explicit means to deal with such a multilevel structure (students in classes and classes in schools), and thus do not account for the resulting variance structures (i.e. students in the same classes tend to be more similar than students in different classes). In order to allow the trees deal with the multilevel structure an aggregation of all variables to the higher levels was performed.
All variables on lower levels (individual, class) were aggregated to each of the higher levels (class, school).
That is, for each variable given on the individual level a mean value on the class level and another on the school level was calculated and added to the data set as additional variable. In the following, the aggregated version of a variable \texttt{var} to class level is denoted by \texttt{var-aggCL}, the aggregated version of a variable to school level by \texttt{var-aggSL}. These aggregated variables where added to the set of variables from which to built the tree models, see Section \ref{sec:classmethods} for further details.

\section{Methods} \label{sec:methods}

\subsection{Classification Models} \label{sec:classmethods}

For the subsequent analysis classification and regression trees (CART, \citealp{Breiman&1984}) were employed. A popular implementation of CART based trees can be found in the \texttt{R} \citep{R2021} package \texttt{rpart} \citep{Therneau&2019}. The CART algorithm constructs a tree by recursive binary partitioning of the complete predictor space into rectangular regions.

In the following some important expressions in the context of growing a tree are briefly explained.
The full data set is called the \textit{root node} of the tree, it is visualized at the top of the tree structure. It is then partitioned by a recursive approach into a set of rectangular regions, called \textit{nodes} of the tree, that are as homogeneous as possible with respect to response.
In the first step the full data is partitioned by a binary decision regarding a certain predictor variable into two rectangular sub-regions called \textit{child nodes}. In subsequent steps each of these two regions can be partitioned further, and this process is repeated within each of the resulting sub-regions until some stopping criterion is fulfilled. To determine the binary partition at a given step in the process it is evaluated which predictor variable and value leads to an optimal partition with regard to a criterion measuring the homogeneity of the response in the resulting child nodes \citep{Breiman&1984, Hastie&2009, Therneau&2019}. The predictor variable and value that are chosen to define the partition are called \textit{split variable} and \textit{split point/value}. As each split variable is chosen to increase the homogeneity of the resulting child nodes as much as possible, the (early) selection of variables for splitting indicates their relevance with respect to response to some extent. A node where no more decision needs to be made is called a \textit{terminal node or leaf}, they are visualized at the end of the branches of the tree. For every node involving a binary decision the chosen split variable and value are provided in the tree visualization. In a classification tree a node also contains information on the observed class frequencies and the resulting class prediction.

In each terminal node a simple model is fit to the respective data in order to predict the response variable. In case of classification trees the response is predicted based on the (absolute or relative) class frequencies in a terminal node: the class prediction is given by the most frequently observed class.

To predict the class of an observation with the grown tree this observation is sent down the appropriate path in the tree: in each node the respective binary condition is checked for the observation at hand, and it is sent either to the left or the right child node depending on whether the binary condition is true or false.

To determine the optimal size of a tree (too large and complex trees might overfit the data, too simple trees might not capture all features) cost-complexity pruning is applied. A large tree is grown which is then pruned back. The degree of pruning depends on the (nonnegative) complexity parameter \texttt{cp} governing the size and complexity of the tree. The value $\texttt{cp}=0$ corresponds to the unpruned tree, increasing values of \texttt{cp} result in smaller and less complex trees. The optimal \texttt{cp} parameter is chosen by cross-validation.
Specifying a value for the argument \texttt{cp} in the \texttt{rpart} package constitutes a form of pre-pruning in order to save computation time by not pursuing splits that are likely to be pruned off within the cross-validation procedure, anyway.

\subsubsection{Issues in the application of trees to pre-processed data} \label{sec:apptrees}

When applying trees to school transition data several issues should be pointed out.

First, an advantage of trees is the possibility to implement a handling strategy for missing values (NA). Several approaches have been introduced (some of them specifically designed for classification trees), for a short overview and comparison see \cite{Twala&2008} and references therein. In this work the approach based on so-called surrogate splits is employed \citep{Breiman&1984, Therneau&2019}. These surrogate splits are used in case the (primary) splitting variable is not available in the observation which is sent down the tree. In that case (one of) the surrogate variables is used in order to send the observation further down the tree until it reaches a terminal node. In contrast, regression models are not able to issue a prediction for a case containing missing values in one or more predictors. Non-complete cases have either be deleted from the data or the missing values have to be imputed.

Second, as mentioned in Section \ref{sec:preproc}, the school transition data (like other educational data) exhibits a multilevel structure.
To account for the different levels in the data within the tree models, the aggregation of lower level variables to all higher levels is proposed, see Section \ref{sec:multilevel} and \ref{sec:classmethods} for details. By letting the trees grow on a data set containing not only the original variables but also their aggregations to higher levels, they are allowed to choose split variables only on a single or on multiple levels. Thus, incorporating the different levels is implicitly done by allowing the trees to decide in a data-driven procedure which of the levels of a variable are relevant.

In order to investigate the usefulness (e.g. with regard to predictive performance) of the proposed approach for trees to deal with the multilevel structure several competitive tree models were considered. They are obtained by letting a tree grow on different data sets containing only variables on individual level (TreeInd), or variables on individual and meta levels (class, school; TreeIndMeta), or individual and/or meta variables jointly with their aggregated versions (TreeIndAgg, TreeIndMetaAgg). This comparison can provide information on whether the aggregated variables are selected at all for splits, and whether they can improve the predictive performance.
In the simulation study described in Section \ref{sec:predperf} these tree models are compared to fixed and mixed effect regression models.

Finally, it should be mentioned that no pre-selection of variables is required for trees, while this might be necessary for classical regression models, where the model building process is often confirmatory.
For the analysis of educational data the variables are usually pre-selected based on known educational theories. However, if data is analyzed for which no theories about the educational background exist, this approach might not be feasible.
Trees on the other hand do not require a pre-selection of variables, they can essentially deal with arbitrary numbers of predictors.
To utilize this advantage for regression models, this work proposes to use trees not only for prediction purposes in the educational context, but also as a heuristic variable selection approach. This procedure is explained in more detail in Section \ref{sec:variablesel}. The pre-selection of variables by the trees is also related to the question whether the chosen split variables are in accordance to existing educational theories, and is discussed in Section \ref{sec:relevancepredictors}.
Recently, an approach to combine trees with a multilevel (generalized linear mixed) model in the terminal nodes of the tree was proposed for a psychological application \citep{Fokkema&2021}. However, the drawback of this approach is that the variables employed for the model in the terminal nodes have to be selected beforehand. In this application, the possibility of using the nonparametric CART trees to perform variable pre-selection will be investigated.

\subsubsection{Comparison to alternative classification models}

In this study classification trees are compared to traditional binary fixed effect regression models (generalized linear models, GLMs), and to binary mixed models (generalized linear mixed models, GLMMs). For an overview on regression models see, for example, \cite{Fahrmeier&2013}.

While (generalized) linear models ignore the above described multilevel structure in the data, (generalized) linear mixed models are designed to model multilevel behaviour explicitly through random effects. In a generalized linear mixed model the linear predictor is extended to contain random effects.
Mixed effect models have become popular in educational sciences, where they are typically referred to as multilevel models \citep{GelmanHill}.
The fixed effect models were fitted with the standard \texttt{glm} function of the \texttt{stats} package in \texttt{R} \citep{R2021},  the mixed effect models within the \texttt{lme4} \texttt{R} package \citep{Bates&2015}.
Models with only random intercepts and models with both, random intercepts and slopes, were investigated for comparison.

\subsection{Assessment of predictive performance} \label{sec:veriscores}

Predictive performance of the models is compared with different verification metrics.

Based on a confusion matrix (see, for example, \citealp{Fawcett2006}) obtained for a fixed classification threshold $\tau$, the true positives (TP), false positives (FP), true negatives (TN), and false negatives (FN) are counted across all prediction cases $i=1,\ldots,n$. Then the classification error rate can be obtained as $\frac{FN + FP}{n}$. Conversely, the rate of correct classifications is given by $\frac{TN + TP}{n}= 1 - \frac{FN + FP}{n}$.

To evaluate the classification performance across all possible classification thresholds $\tau$, the respective Receiving Operator Characteristic (ROC, see, for example, \citealp{Hastie&2009, Fawcett2006}) curves are inspected. An ROC curve is obtained by plotting the true positive rate (sensitivity) versus the true negative rate (specificity) of the classifier as a function of $\tau \in [0,1]$.
The area under the curve (AUC) has a value between 0.5 and 1, where values close to 1 indicate high classification performance.

Furthermore, the quality of the estimated class probabilities is assessed by the Brier score \citep{Brier1950}. This score assigns a numerical value to a pair $(\hat{p},y)$, where $\hat{p}$ is a predicted probability (probabilistic forecast).
For the special case of a binary response $y_i \in \{0,1\}$ with predicted probabilities $\hat{p}_{i1}=\hat{p}_i$ for class 1 (success class) a simplified version of the Brier score is given by:
$\textup{BS}=\frac{1}{n} \, \sum_{i=1}^n \; (\hat{p}_i - y_i)^2,$
where the individual values are averaged over all prediction cases $i=1,\ldots,n$.

\section{Application} \label{sec:casestudy}

In the following school transition rates in Austria are predicted based on the Austrian educational standards data. The results are presented according to the introduced issues (i), (ii), and (iii), see Section \ref{sec:Intro}.
To illustrate how to apply classification trees to educational data a file with example R code on data pre-processing, tree growing and visualization is provided as supplementary material to this paper. As the data set used for the study presented here is not publicly available, the R code example is based on another data set. This public data set contains similar variables to the ones used in our analysis, however it includes no variables on the class level.

\subsection{Relevance of predictors} \label{sec:relevancepredictors}

The first question to be investigated is whether the trees select splitting variables that are in accordance with existing theories from educational sciences and can be of help in gaining deeper understanding of the relationships.

To that end, the variables chosen by the trees for splits as well as their importance in the tree growing process are analyzed.
Variables chosen in the very first splits can be considered as informative for class membership. For the data analyzed here respective approaches on which variables are most informative for predicting school transition are readily available (see, for example, \citealp{Flores&2011}) and can be compared to the trees choices.

\subsubsection{Example tree grown on full data set}

\begin{figure}
\centering
\includegraphics[width=145mm, height=85mm]{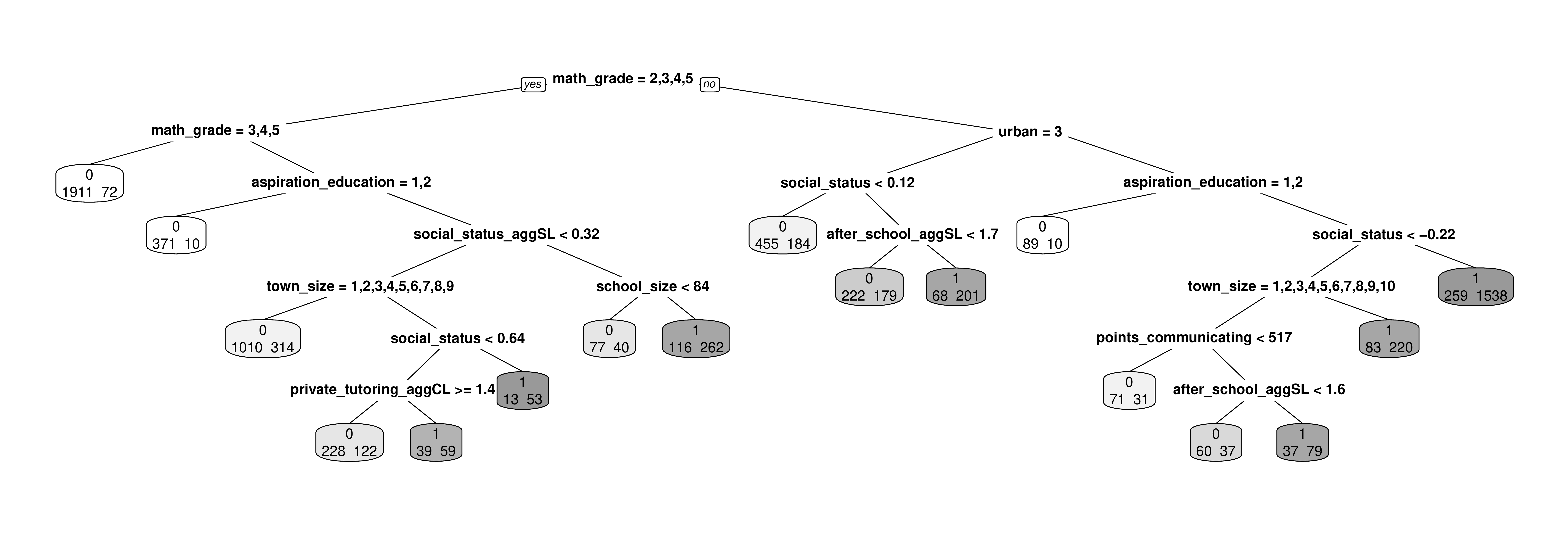}
\caption{A classification tree based on individual variables, variables on class and school level, as well as aggregated variables to both meta levels, grown with $\texttt{cp}=0.004$, plotted with \texttt{rpart.plot} \citep{Milborrow2019}.} \label{ExampleTree}
\end{figure}

Figure \ref{ExampleTree} shows a classification tree grown on the full data set containing the individual variables, the meta variables on class and school level, as well as all variables aggregated from the lower levels to each of the higher levels. The tuning parameter  \texttt{cp} in the example tree was chosen as $\texttt{cp}=0.004$ (for details see Section \ref{sec:predperf}).

When looking at the first splits in the left direction, it becomes clear, that especially for high values of \texttt{math-grade} (low grades in math) in combination with small values of \texttt{aspiration-education} (aspiration of highest educational degree of parents for their children is low) and with small sized municipalities (low values of \texttt{town-size} corresponding to rural areas) there is a strong indication that the respective student will not make the transition to AHS. The variables \texttt{math-grade}, \texttt{education-aspiration}, and \texttt{town-size} seem to lead to very clear separation between the two classes.
However, in view of Table \ref{ConfusionMatrixTreeIndMetaAgg} and the resulting proportions of correctly predicted \texttt{0} and \texttt{1} cases it becomes also evident, that it is slightly easier to predict class \texttt{0} (student will not attend AHS), than predicting class \texttt{1} (student will attend AHS). This can partially be explained by the fact, that there is a class imbalance in the data set (approximately 5,100 cases of \texttt{0}, and approximately 3,400 cases of \texttt{1}), meaning a relatively small portion of students is attending AHS at all.

The fact that prediction of class \texttt{1} is somewhat more difficult is supported further by educational theories, which state that the set of factors leading a student to visit AHS is much more complex than the set of factors leading a student not to visit AHS (see, for example, \citealp{Bruneforth2015}).

\subsubsection{Interpretation of split variables in educational context}

The variables chosen for splits in the tree in Figure \ref{ExampleTree} are in accordance with the described educational approach. For illustration purposes a specific path in the tree is considered, namely the path starting at the top and following the directions left, right, right, left, left (in that order). It describes students with grade 2 in math and parents having a higher
educational attainment for them, embedded in a school context in which most parents have a medium to low social status, and attending school in small to medium villages.
These students are mostly classified not to attend AHS (class 0) in the next year. According to educational theory context factors such as a lower education and social status of the parents decrease the student's chance of attending a higher school track \citep{Bruneforth2015}. For Austria, it is also known that less students from smaller villages actually attend AHS in comparison to students living in urban areas. One reason is that attending the nearest AHS often requires a lot of travel time or even relocation for students from smaller villages \citep{Mayrhofer&2019}. In contrast, the parents educational aspiration for their children has a positive impact on selecting a higher school track \citep{Astleithner&2021}.

The choice of the variables \texttt{private-tutoring-aggCL} and \texttt{after-school-aggSL} for splitting is surprising from an educational perspective but not contradicting established theories.
The variable \texttt{private-tutoring-aggCL} implies that students of a class having a higher average of hours for private tutoring are less likely to visit AHS. Thus, if students of a class tend to require more hours of tutoring that may be an indication for low teaching quality.
The variable \texttt{after-school-aggSL} indicates that students at a school having a higher average of hours spent in after-school supervision are more likely to visit AHS. This could suggest that open forms of all-day schools are beneficial for the students educational forthcoming.

\subsubsection{Frequency of variables chosen for splits as measure of importance}

Table \ref{CountsSplitVariables} presents for a subset of variables how often they were chosen as split variables across the 50 trees grown on the 50 randomly sampled training sets. To be more specific, it was counted how often a certain variable occurs within each individual tree, and these individual counts of the 50 trees were summed up to obtain an overall count across all trees.

For comparison these counts are presented for two different tree models, which result from growing a tree based on individual variables and meta variables (TreeIndMeta), and a tree based on individual and meta variables, as well as their aggregations (TreeIndMetaAgg).
For details see Section \ref{sec:variablesel}, where Table \ref{ModelAbbreviations} also introduces the abbreviations of the models.

For both tree models essentially the same variables were selected in the growing process, with similar frequencies. However, the tree model IndMetaAgg that was allowed to choose also from aggregated variables, indeed chose an aggregated variable for splits quite frequently. If measuring the relevance of a variable for predicting the school transition by the (absolute) frequency this variable was chosen for splits, both trees considered here indicate that the most important variables are \texttt{math-grade}, \texttt{aspiration-education}, and \texttt{social-status}.

\begin{table}
\caption{Counts of how often variables where chosen as split variables across the 50 trees from the repetitions.}
{\begin{tabular}{lcc} \toprule
 & \multicolumn{2}{c}{Tree Model} \\ \cmidrule{2-3}
 Variable & TreeIndMetaAgg & TreeIndMeta \\ \midrule
  \texttt{after-school-aggSL} & 63 & - \\
 \texttt{aspiration-education} & 124 & 134 \\
 \texttt{math-grade} & 100 & 100 \\
 \texttt{social-status} & 117 & 131 \\
 \texttt{town-size} & 49 & 98 \\
 \texttt{urban} & 46 & 50 \\
 \texttt{federal-state} & 43 & 94 \\
 \texttt{points-calculating} & 14 & 23 \\ \bottomrule
\end{tabular}}
\label{CountsSplitVariables}
\end{table}

To further investigate the importance of the variables, the variables appearing specifically in the first, second and third split of the 50 trees were investigated. Here, it was found that \texttt{math-grade} is always selected as the fist and the second split variable - in each of the 50 trees, for both tree models presented in Table \ref{CountsSplitVariables}.
The third split showed a bit more variation, but nonetheless confirms the observations in Table \ref{CountsSplitVariables}.
For the tree in the first column, the variables chosen for the third split in the 50 grown trees were \texttt{urban} (28 times), \texttt{after-school-aggSL} (14 times), \texttt{social-status} (7 times), and \texttt{aspiration-education} (1 times). For the tree in the second column, the variables \texttt{urban} (35 times), \texttt{social-status} (11 times), \texttt{aspiration-education} (3 times), and \texttt{town-size} (1 times) where chosen in the third splits of the 50 trees.
This indicates, that besides \texttt{math-grade}, the variables \texttt{urban} and \texttt{social-status} are of high importance to predict the school transition.
As explained earlier in this section these variable choices are highly consistent with educational theory.

\subsection{Predictor selection for (regression) models} \label{sec:variablesel}

This section discusses the question whether the variables selected by trees in the growing process are not only useful from an interpretational point of view (see Section \ref{sec:relevancepredictors}), but can also be utilized to built reasonable regression models when either the number of predictors to choose from is very large, or if no educational theory exists that can help to reasonably select important predictors.

For new types of variables e.g. coming from process data in large-scale assessments (see, for example, \citealp{Salles&2020}) where little is known about the relationships of the variables, a pre-selection can be difficult, see Section \ref{sec:outlook}.

This work presents a data-driven (heuristic) approach of variable pre-selection based on a tree.
For this procedure a reasonably complex tree (see Section \ref{sec:predperf}) is grown on the full data set containing the original variables and possibly the aggregated variables. An example of such a tree with $\texttt{cp}=0.004$ can be found in Figure \ref{ExampleTree}.

In order to investigate the relevance of the different levels (individual, class, school) tree models grown on two different sets of variables (TreeIndAgg and TreeIndMetaAgg) were used for predictor selection.
As mentioned before, the tree models do not necessarily include all variables on all levels, but choose in a data-driven way which of these effects are most informative with regard to classification. The variables selected by the two types of trees were used to construct corresponding regression models: fixed effect regression models (GLMInd, GLMIndMeta), random intercept models (GLMMInd.I, GLMMIndMeta.I), and also random intercept random slope models (GLMMInd.S, GLMMIndMeta.S). Random intercepts were generally introduced on class as well as on school level, while random slopes were introduced for variables for which their corresponding aggregated version had been selected. The tree models, their corresponding regression models, and the associated types of variables are specified in Table \ref{ModelAbbreviations}.

In addition, two other tree models were included into the model comparison, one model based only on individual variables (TreeInd), and one model based on individual, class, and school variables (TreeIndMeta). These trees were grown on sets of variables not containing any of the aggregated variables, to investigate their possible effect on the performance.

\begin{table}
\caption{Overview on considered tree models, their abbreviations, and corresponding regression models.}
{\begin{tabular}{ll} \toprule
 \textbf{Tree Model} & \textbf{Corresponding Regression Model} \\ \midrule \midrule
\begin{minipage}{0.5\textwidth}
\textbf{TreeInd}\\
\textit{Based on individual variables}\end{minipage} &
\begin{minipage}{0.4\textwidth}~\end{minipage}\\
\midrule
\begin{minipage}{0.5\textwidth}
\textbf{TreeIndAgg}\\
\textit{Based on individual variables, and
their aggregations to class and school level}
\end{minipage} & \begin{minipage}{0.4\textwidth} \text{GLMInd}, \text{GLMMInd.I}, \text{GLMMInd.S}\end{minipage} \\
\midrule
\begin{minipage}{0.5\textwidth}
\textbf{TreeIndMeta}\\
\textit{Based on individual, class and school variables}\end{minipage} &
\begin{minipage}{0.4\textwidth}~\end{minipage}  \\
\midrule
\begin{minipage}{0.5\textwidth}
\textbf{TreeIndMetaAgg}\\
\textit{Based on individual, class and school variables,
and their aggregations to class and school level}\end{minipage} &
\begin{minipage}{0.4\textwidth}
\text{GLMIndMeta}, \text{GLMMIndMeta.I}, \text{GLMMIndMeta.S}
\end{minipage}\\
\midrule
\begin{minipage}{0.5\textwidth}
\textbf{No corresponding tree}\\
\textit{Based on variables chosen from
educational theory}\end{minipage} & \begin{minipage}{0.4\textwidth}
 \text{GLMedu}, \text{GLMMedu.I}\end{minipage} \\
\midrule \midrule
\multicolumn{2}{l}{
\begin{minipage}{0.9\textwidth}
\textit{Note}: GLM refers to a fixed effect logistic model, GLMM to a mixed effect logistic model,
the letter .I denotes a mixed effect model with only random intercepts,
the letter .S a mixed effect model with random intercepts and random slopes.
\end{minipage}
}\\
\bottomrule
\end{tabular}}
\label{ModelAbbreviations}
\end{table}

To have a comparison of regression models based on the tree selected variables with models that rely solely on the educational approach by  \cite{Flores&2011}, two further regression models (GLM.edu, GLMMedu.I) were considered. According to this theory the student's choice of the next school she/he will be attending depends on the following variables: gender, educational level of mother and father, number of books at home, existence of computer, study place and internet connection at home, the level of mathematical and linguistic communication competence and the parents educational attainment for their children. With exception of the three variables describing the student's equipment at home these variables are available in the data set at hand and were used as predictors in the confirmatory ``educational'' regression models.

\subsection{Results for predictive performance} \label{sec:predperf}

In this section the predictive performance of the different models is investigated to see whether trees are comparative to classical regression models.

A short preliminary study was conducted in order to determine optimal values for the tuning parameters of the trees, specifically for
the complexity parameter \texttt{cp} in the \texttt{rpart} function, see Section \ref{sec:classmethods}.
A preliminary analysis for the given data indicated that quite small values of \texttt{cp} are required in order to obtain reasonably complex trees.
A sequence of \texttt{cp} values between 0 and 0.1 (in 0.001 steps) was investigated in more detail. This study indicated that the optimal value of \texttt{cp} in terms of the error rate is between 0.003 and 0.005. For values smaller than 0.003 and larger than 0.005 the performance deteriorates monotonically.
With regard to the use of the trees for variable selection (see Section \ref{sec:variablesel}), a compromise between a more complex and a simpler tree was chosen, thus the parameter \texttt{cp} was set to 0.004 for all analyses involving a tree.
For the remaining tuning parameters of the \texttt{rpart} function the default values for classification where used, as preliminary studies suggested them to be appropriate.

\begin{figure}
\centering
\includegraphics[width=140mm]{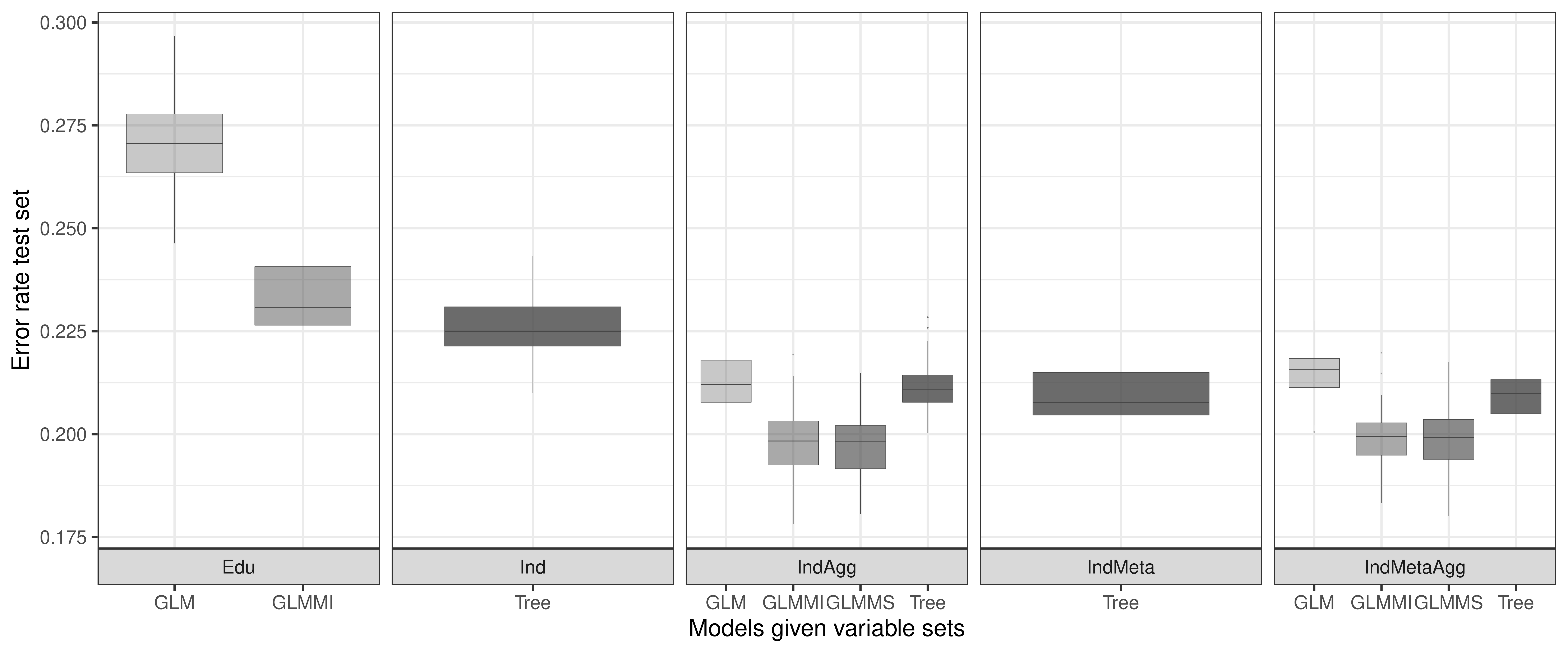}
\caption{Boxplots of error rates of the considered classification models (GLM, GLMM.I, GLMM.S, Tree), grouped by the variable sets (Edu, Ind, IndAgg, IndMeta, IndMetaAgg), based on 50 repetitions.} \label{predPerfModels}
\end{figure}

The predictive performance of the tree and regression models is compared in a simulation study, where the full data set with 8,520 cases was randomly partitioned into a training set of size 5,000 and a test set of size 3,520. This procedure was repeated 50 times. The models were fit on each of the 50 training sets, and the response was predicted on each of the 50 test sets, yielding the test set error rate and Brier score.

\begin{figure}
\centering
\includegraphics[width=140mm]{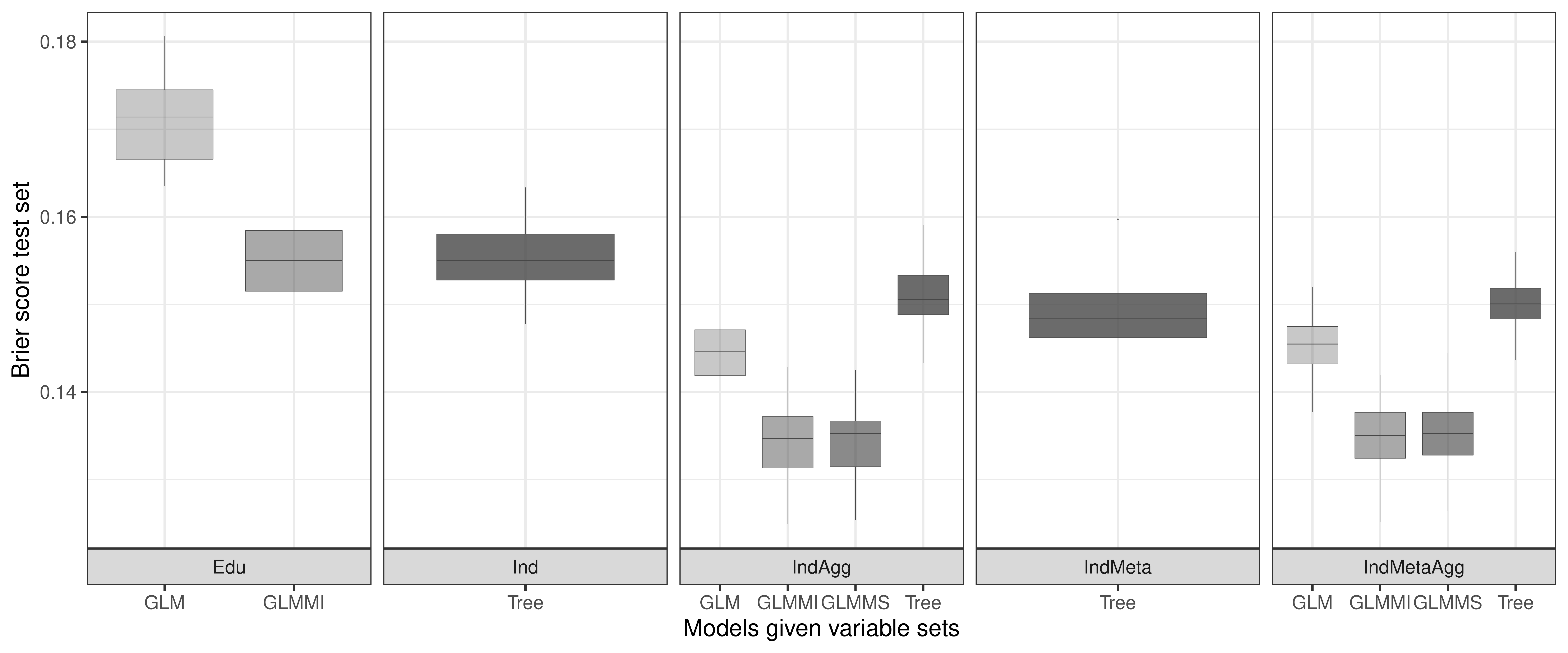}
\caption{Boxplots of Brier scores of the considered classification models (GLM, GLMM.I, GLMM.S, Tree), grouped by the variable sets (Edu, Ind, IndAgg, IndMeta, IndMetaAgg), based on 50 repetitions.} \label{predPerfModels1}
\end{figure}

Figure \ref{predPerfModels} shows boxplots of the error rates, and Figure \ref{predPerfModels1} boxplots of the Brier scores for all models on the 50 test sets. The best performance with respect to both verification measures is achieved by the mixed models based on variables pre-selected by trees grown on the variable set IndAgg or IndMetaAgg (see Table \ref{ModelAbbreviations}). All models (trees and regression models) based on the variable set IndMegaAgg exhibit a slightly better performance than their corresponding models based on the variable set IndAgg. In terms of error rate the trees based on the variable sets IndAgg and IndMetaAgg even slightly outperform the corresponding fixed effect regression models.

The best models among the trees in terms of both scores are the ones based on the variable sets IndAgg, IndMet and IndMetaAgg. They show improved performance over a tree based only on individual variables (variable set Ind). This implies that using variables only on an individual (student) level might not be sufficient, and adding variables on the meta levels class and school (in their original form or by aggregating to that level) can further improve classification performance.

In addition, the fixed effect as well as the mixed effect models based on the variables pre-selected by the trees clearly improve performance over the respective fixed and mixed model based on the variables selected by educational expert knowledge. This suggests further usefulness of trees for understanding, modelling and predicting educational variables. The trees are able to pre-select an informative set of variables for building regression models, and to improve performance over models using predictors selected in literature reviews by  educational experts.

Overall, the considered models show a high classification accuracy, as the majority of error rates are between 18 \% and 25 \% (with the GLM.edu model being an exception), thus yielding an accuracy of around 80 \%.

\begin{figure}
\centering
\includegraphics[width=135mm]{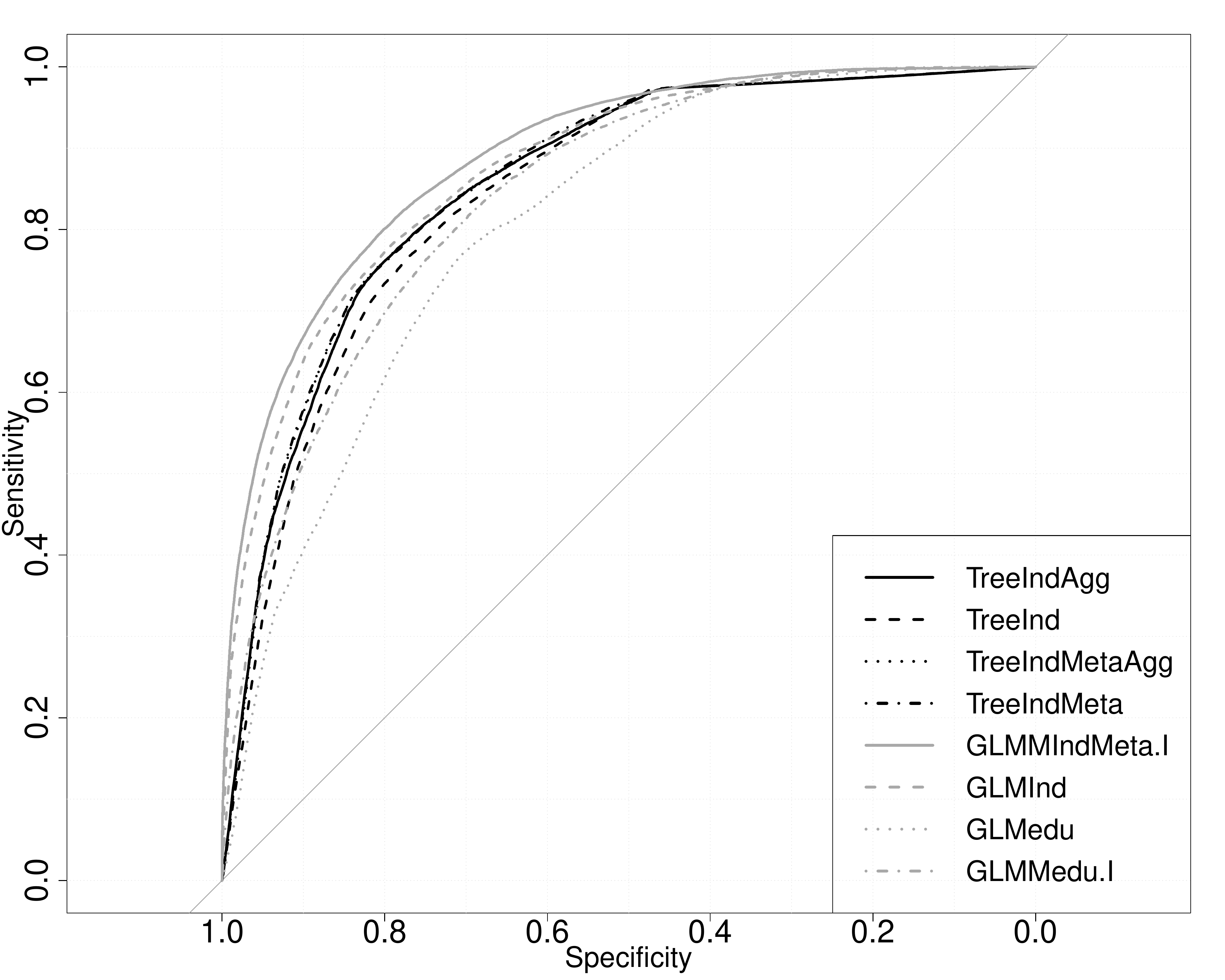}
\caption{ROC curves of classification models (GLM, GLMM.I, GLMM.S, Tree) for the different variable sets (Edu, Ind, IndAgg, IndMeta, IndMetaAgg), based on aggregation of ROC values over 50 repetitions.} \label{RocCurvesModels}
\end{figure}

To investigate the classification performance in more detail, ROC curves for a subset of the considered models are presented in Figure \ref{RocCurvesModels}. Most ROC curves are quite similar, the area under the curves (AUC) of the models ranges between 0.8 and 0.89, indicating high classification accuracy for all models. The best ROC curve belongs to the mixed model with random intercepts based on variables pre-selected from the set IndMeta, closely followed by the trees grown on the variable sets IndAgg, IndMeta, and IndMetaAgg, whose curves are nearly identical. The least classification performance is exhibited by the fixed effect model based on the educational variables.

It is also of interest whether both classes can be predicted with equal accuracy. In order to investigate this each of the 50 trees grown on one of the 50 training sets was utilized to obtain class predictions on the respective test set. The predicted classes were compared with the true classes observed in the test set. A confusion matrix of the true classes and the predicted classes was built, aggregating the cross-counts over all predictions on the 50 test sets. As an example the result for the tree model TreeIndMetaAgg is presented in Table \ref{ConfusionMatrixTreeIndMetaAgg}.

\begin{table}
\caption{Classes predicted by TreeIndMetaAgg, aggregated over 50 repetitions.}
{\begin{tabular}{lccc} \toprule
 & \multicolumn{2}{c}{Class Prediction} &  \\ \cmidrule{2-3}
 True Class & \texttt{0} & \texttt{1} & Total \\ \midrule
 \texttt{0} & 88179 & 17316 & 105495 \\
 \texttt{1} & 19552 & 50953 & 70505 \\ \bottomrule
\end{tabular}}
\label{ConfusionMatrixTreeIndMetaAgg}
\end{table}

Class \texttt{0} (students not attending AHS in the next year) is correctly predicted much more frequently than class \texttt{1}. To be more specific, in around 84\% of the overall cases of \texttt{0} the model correctly predicts this class, while in around 72\% of the overall cases of \texttt{1}, this class is correctly predicted, indicating a slightly higher specificity of the model than sensitivity (when aggregating the cases over the 50 repetitions). This phenomenon was already discussed as part of the interpretation of the tree in Figure \ref{ExampleTree}.

\section{Discussion and outlook} \label{sec:outlook}

In this work classification trees are employed to predict school transition rates in Austria.
So far the use of machine learning based methods is rare in the area of educational sciences. Therefore, this work aims to further
investigate the benefits of applying these methods to discover structures in secondary analyses of educational data. Several important issues were raised within this study. The first issue (i) investigated the usefulness of the variables chosen for splits in the trees for educational theories. It turned out that the variables selected by the trees are in strong accordance with existing approaches in education explaining school transition.
The application revealed that the trees (also) select variables which are not commonly used, thus providing potential starting points for refining and extending existing approaches.
This could be of specific help in approaches, which do not only contain direct effects on school transition, but
describe a far more complex picture with underlying direct and indirect mechanisms of
action, see for example \cite{Maaz&2006} or \cite{Klinge2016}.

The second issue (ii) investigated whether trees can be utilized as a data-driven pre-selection procedure for variables used e.g. in regression models. This question is directly related to the last issue (iii) which investigated whether the predictive performance of trees and regression models are comparable.
A simulation study to compare predictive performance also revealed that the trees show comparable predictive performance to the classical regression models. They outperformed regression models using only predictors chosen from educational theory, and show at least comparable performance to some of the regression models using tree-based variable pre-selection.
These findings suggest that a tree-based variable selection for regression models to analyze educational data could be quite beneficial. This issue will be investigated further in future research, where a more standardized or automatic tree-based selection procedure may be developed.

A further advantage of the trees is that they can explicitly deal with missing values, while (generalized) linear (mixed) models cannot predict cases where one or more observations of the predictors are missing. For data from educational sciences, the missing values are often imputed (see, for example, \citealp{vonDavier&2009}). Future research may also involve comparisons of different imputation methods for trees (as in \citealp{Twala&2008}) to regression models with and without imputation of missing values.

Note that in this analysis the students' competencies are represented by weighted Likelihood estimates (WLEs). Plausible values (PVs) were not used since the main objective was not interpretation of regression parameters based on unbiased estimates but variable selection and comparison of tree based classification with standard models. In this comparison all models were based on the very same WLEs.

The proposed pre-selection procedure of variables for regression models will be particularly useful in educational scenarios for which theories about the dependence between variables do not yet exist. For example, one may apply this procedure for analyzing new forms of process data in large-scale assessments such as PISA \citep{Gobert&2015, LezhninaKismihok2021}.

\section*{Notes}

The data was provided by the research data library of the Federal Institute for Quality Assurance of the Austrian School System (https://iqs.gv.at/fdb).


\begin{thebibliography}{}

\bibitem[Astleithner et~al., 2021]{Astleithner&2021}
Astleithner, F., Vogl, S., and Parzer, M. (2021).
\newblock Zwischen {W}unsch und {W}irklichkeit: {Z}um {Z}usammenhang von
  sozialer {H}erkunft, {M}igration und {B}ildungsaspirationen.
\newblock {\em \"{O}sterreichische Zeitschrift f\"{u}r Soziologie},
  46:233--256.

\bibitem[Bates et~al., 2015]{Bates&2015}
Bates, D., M\"{a}chler, M., Bolker, B., and Walker, S. (2015).
\newblock Fitting {L}inear {M}ixed-{E}ffects {M}odels {U}sing {lme4}.
\newblock {\em Journal of Statistical Software}, 67:1--48.

\bibitem[BIFIE, 2018]{BIFIE2018}
BIFIE (2018).
\newblock {\em Context questionnaires of educational standards test 2018 in
  mathematics for grade four}.
\newblock Bundesinstitut BIFIE Austria.

\bibitem[BIFIE, 2019]{BIFIE2019}
BIFIE (2019).
\newblock {\em Standard\"{u}berpr\"{u}fung 2018. Mathematik, 4. Schulstufe.
  Bundesergebnisbericht.}
\newblock Bundesinstitut BIFIE Austria.

\bibitem[Breiman et~al., 1984]{Breiman&1984}
Breiman, L., Friedman, J.~H., Olsen, R.~A., and Stone, C.~J. (1984).
\newblock {\em Classification and {R}egression {T}rees}.
\newblock Chapman \& Hall/CRC.

\bibitem[Brier, 1950]{Brier1950}
Brier, G.~W. (1950).
\newblock Verification of forecasts expressed in terms of probability.
\newblock {\em Monthly Weather Review}, 78:1--3.

\bibitem[Bruneforth and Itzlinger-Bruneforth, 2015]{Bruneforth2015}
Bruneforth, M. and Itzlinger-Bruneforth, U. (2015).
\newblock Die {S}chulwahl von {S}ch\"{u}ler/innen am {E}nde der 8. {S}chulstufe
  im {L}ichte ihrer {M}athematikkompetenz.
\newblock In Stock, M., Schl\"{o}gl, P., Schmid, K., and Moser, D., editors,
  {\em Kompetent – wof\"{u}r? Tagungsband zur 4. \"{O}sterreichischen
  Konferenz f\"{u}r Berufsbildungsforschung am 3./4. Juli 2014}, pages
  263--282. StudienVerlag Innsbruck.

\bibitem[Fahrmeir et~al., 2013]{Fahrmeier&2013}
Fahrmeir, L., Kneib, T., Lang, S., and Marx, B. (2013).
\newblock {\em Regression: {M}odels, {M}ethods and {A}pplications}.
\newblock Springer Berlin.

\bibitem[Fawcett, 2006]{Fawcett2006}
Fawcett, T. (2006).
\newblock An {I}ntroduction to {ROC} {A}nalysis.
\newblock {\em Pattern Recognition Letters}, 27:861--874.

\bibitem[Fokkema et~al., 2021]{Fokkema&2021}
Fokkema, M., Erdbrooke-Childs, J., and Wolpert, M. (2021).
\newblock Generalized linear mixed-model ({GLMM}) trees: {A} flexible
  decision-tree method for multilevel and longitudinal data.
\newblock {\em Psychotherapy Research}, 31:329--341.

\bibitem[Gao and Rogers, 2011]{GaoRogers2011}
Gao, L. and Rogers, W.~T. (2011).
\newblock Use of tree-based regression in the analyses of {L2} reading test
  items.
\newblock {\em Language Testing}, 28:77--104.

\bibitem[Gelman and Hill, 2007]{GelmanHill}
Gelman, A. and Hill, J. (2007).
\newblock {\em Data {A}nalysis {U}sing {R}egression and
  {M}ultilevel/{H}ierarchical {M}odels}.
\newblock Analytical Methods for Social Research. Cambridge University Press.

\bibitem[Gil-Flores et~al., 2011]{Flores&2011}
Gil-Flores, J., Padilla-Carmona, M.~T., and Su{\'{a}}rez-Ortega, M. (2011).
\newblock Influence of gender, educational attainment and family environment on
  the educational aspirations of secondary school students.
\newblock {\em Educational Review}, 63:345--363.

\bibitem[Gobert et~al., 2015]{Gobert&2015}
Gobert, J.~D., Kim, Y.~J., Sao~Pedro, M.~A., Kennedy, M., and Betts, C.~G.
  (2015).
\newblock Using educational data mining to assess students' skills at designing
  and conducting experiments within a complex systems microworld.
\newblock {\em Thinking Skills and Creativity}, 18:81--90.

\bibitem[Gro{\ss} et~al., 2016]{Gross&2016}
Gro{\ss}, J., Robitzsch, A., and George, A.~C. (2016).
\newblock Cognitive diagnosis models for baseline testing of educational
  standards in math.
\newblock {\em Journal of Applied Statistics}, 43:229--243.

\bibitem[Hastie et~al., 2009]{Hastie&2009}
Hastie, T., Tibshirani, R., and Friedman, J. (2009).
\newblock {\em The {E}lements of {S}tatistical {L}earning: {D}ata {M}ining,
  {I}nference, and {P}rediction}.
\newblock Springer Series in Statistics. Springer New York, 2 edition.

\bibitem[Klinge, 2016]{Klinge2016}
Klinge, D. (2016).
\newblock {\em Die elterliche \"{U}bergangsentscheidung nach der Grundschule.
  Werte, Erwartungen und Orientierungen.}
\newblock Springer VS Wiesbaden.

\bibitem[Lezhnina and Kismih\'{o}k, 2021]{LezhninaKismihok2021}
Lezhnina, O. and Kismih\'{o}k, G. (2021).
\newblock Combining statistical and machine learning methods to explore
  {G}erman students' attitudes towards {ICT} in {PISA}.
\newblock {\em International Journal of Research \& Method in Education},
  45:180--199.

\bibitem[Maaz et~al., 2006]{Maaz&2006}
Maaz, K., Hausen, C., McElvany, N., and Baumert, J. (2006).
\newblock Stichwort: {\"{U}}berg\"{a}nge im {B}ildungssystem.
\newblock {\em Zeitschrift f\"{u}r Erziehungswissenschaft}, 9:299--327.

\bibitem[Mayrhofer et~al., 2019]{Mayrhofer&2019}
Mayrhofer, L., Oberwimmer, K., Toferer, B., Neubacher, M., Freunberger, R.,
  Vogtenhuber, S., and Baumegger, D. (2019).
\newblock Indikatoren {C}: {P}rozesse des {S}chulsystems.
\newblock In Oberwimmer, K., Vogtenhuber, S., Lassnigg, L., and Schreiner, C.,
  editors, {\em Nationaler Bildungsbericht \"{O}sterreich 2018. Band 1: Das
  Schulsystem im Spiegel von Daten und Indikatoren}. Leykam Graz.

\bibitem[Milborrow, 2019]{Milborrow2019}
Milborrow, S. (2019).
\newblock {\em rpart.plot: {P}lot 'rpart' {M}odels: {A}n {E}nhanced {V}ersion
  of 'plot.rpart'}.
\newblock R package version 3.0.8.

\bibitem[OECD, 2010]{OECD2010}
OECD (2010).
\newblock {\em Education at a glance, 2010: {OECD} indicators}.
\newblock Organisation for Economic Cooperation and Development (OECD).

\bibitem[{R Core Team}, 2021]{R2021}
{R Core Team} (2021).
\newblock {\em R: A Language and Environment for Statistical Computing}.
\newblock R Foundation for Statistical Computing, Vienna, Austria.

\bibitem[Salles et~al., 2020]{Salles&2020}
Salles, F., Dos~Santos, R., and Keskpaik, S. (2020).
\newblock When didactics meet data science: process data analysis in
  large-scale mathematics assessment in {F}rance.
\newblock {\em Large-scale Assessments in Education}, 8:1--20.

\bibitem[Sinharay, 2016]{Sinharay2016}
Sinharay, S. (2016).
\newblock An {NCME} {I}nstructional {M}odule on {D}ata {M}ining {M}ethods for
  {C}lassification and {R}egression.
\newblock {\em Educational Measurement: Issues and Practice}, 35:38--54.

\bibitem[Therneau and Atkinson, 2019]{Therneau&2019}
Therneau, T. and Atkinson, B. (2019).
\newblock {\em rpart: {R}ecursive {P}artitioning and {R}egression {T}rees}.
\newblock R package version 4.1-15.

\bibitem[Twala et~al., 2008]{Twala&2008}
Twala, B.~E.~T.~H., Jones, M.~C., and Hand, D.~J. (2008).
\newblock Good methods for coping with missing data in decision trees.
\newblock {\em Pattern Recognition Letters}, 29:950--956.

\bibitem[von Davier et~al., 2009]{vonDavier&2009}
von Davier, M., Gonzalez, E., and Mislevy, R. (2009).
\newblock Plausible values: {W}hat are they and why do we need them?
\newblock {\em IERI Monograph Series: Issues and Methodologies in Large-Scale
  Assessments}, 2:9--36.

\end{thebibliography}
\end{document}